\documentclass[reprint,superscriptaddress,amsmath,amssymb,aps,nofootinbib,prx]{revtex4-2}
\usepackage{amsmath}
\usepackage[shortlabels]{enumitem}
\usepackage{amsfonts}
\usepackage{graphicx}
\usepackage{bbold}
\usepackage{color}
\usepackage{hyperref}
\usepackage{verbatim}
\usepackage{cases}
\usepackage{amsfonts}
\usepackage{amssymb}
\usepackage[table]{xcolor}
\usepackage{epsfig}
\usepackage{bm}
\usepackage{setspace}
\usepackage{enumerate}
\usepackage{dcolumn}
\usepackage{bm}
\usepackage{enumitem}
\usepackage{multirow}
\usepackage{multirow, array} 
\usepackage{graphicx}
\usepackage{dcolumn}
\usepackage{bm}
\usepackage{color}
\usepackage{amssymb}
\usepackage{lipsum}
\usepackage{amsmath}
\usepackage{graphicx}
\usepackage{array}
\usepackage{multirow}
\usepackage{booktabs}

\begin{document}

\preprint{APS/123-QED}

\title{Classical harmonic three-body system: An experimental electronic realization}
\author{A. M. Escobar-Ruiz}%
\affiliation{Departamento de F\'{i}sica, Universidad Aut\'onoma Metropolitana Unidad Iztapalapa, San Rafael Atlixco 186, 09340 Cd. Mx., M\'exico}
\thanks{These authors contributed equally to this work.}%
\author{M. A. Quiroz-Juarez}%
\affiliation{Centro de F\'isica Aplicada y Tecnolog\'ia Avanzada, Universidad Nacional Aut\'onoma de M\'exico, Boulevard Juriquilla 3001, Juriquilla, 76230 Quer\'etaro, M\'exico}
\thanks{These authors contributed equally to this work.}%


\author{J. L. Del Rio-Correa}%
\author{N. Aquino}%

\affiliation{Departamento de F\'{i}sica, Universidad Aut\'onoma Metropolitana Unidad Iztapalapa, San Rafael Atlixco 186, 09340 Cd. Mx., M\'exico}



\date{\today}

\begin{abstract}

The classical three-body harmonic system in $\mathbb{R}^d$ ($d>1$) with finite rest lengths and zero total angular momentum $L=0$ is considered. This model describes the dynamics of the $L=0$ near-equilibrium configurations of three point masses $(m_1,m_2,m_3)$ with arbitrary pairwise potential $V(r_{ij})$ that solely depends on the relative distances between bodies. It exhibits an interesting mixed regular and chaotic dynamics as a function of the energy and the system parameters. The corresponding harmonic quantum system plays a fundamental role in atomic and molecular physics. In this work we report on a novel electronic experimental realization of the model as a complementary tool to analyze the rich dynamics of the classical system. Our setup allows us to experimentally explore different regions of behavior due to the fact that the intrinsic parameters and initial states of the system are independently set by voltage inputs. Chaotic and periodic motions are characterized employing time series, phase planes, and the largest Lyapunov exponents as a function of the energy and system parameters. The results show an excellent qualitative as well as quantitative agreement between theory and experiment. 

\end{abstract}

 \maketitle



\section{Introducción} 
The three-body harmonic oscillator (TBHO) in $\mathbb{R}^d$ ($d>1$) with finite rest lengths is a 9-parametric system depending on three arbitrary masses, three rest lengths and three spring constants. For generic pairwise potentials, $V=U(r_{12}) +U(r_{13}) +U(r_{23})$, that only depend on the three relative mutual distances $r_{ij}=\mid {\bf r}_i-{\bf r}_j\mid$ between bodies, this model describes the dynamics near the equilibrium configurations and as such it possesses a wide range of relevant problems it can be applied to both in classical \cite{musielak2014three, PhysRevLett.70.3675} and quantum mechanics \cite{PhysRevLett.112.173001}. In the simplest planar case ($d=2$) with equal masses and equal spring constants the classical system already displays a very rich dynamics for different values of the energy where a power-law statistics that fits the Levi-walk model \cite{RevModPhys.87.483} emerges \cite{PhysRevLett.122.024102,PhysRevE.101.032211}. Even on the invariant manifold of zero total angular momentum, regions of regular and chaotic dynamics coexist in the space of parameters due to the intrinsic nonlinearities induced by non-zero rest lengths. If these rest lengths are set equal to zero the system becomes integrable \cite{Castro1993ExactSF}. Not surprisingly, at zero rest lengths the corresponding quantum system, restricted to the subspace of zero total angular momentum, turns out to be an exactly-solvable problem with spectra linear in quantum numbers \cite{Turbiner_2020}. In variables $\rho_{ij}=r_{ij}^2$, the relevant Hamiltonian operator possesses a hidden algebra  $s\ell(4)$ and it admits a quasi-exactly solvable extension. However, for non zero rest lengths a single exact solution to the Schrodinger equation has not been found so far. Hence, the TBHO also represents a realistic model to test theoretical and numerical tools to elucidate the connection between classical and quantum mechanics in chaotic systems \cite{Gutzwiller1971PeriodicOA}.

In this Letter, we report the physical realization of the classical TBHO using analog electrical components where the basic building blocks are non linearly coupled electrical inductance-capacitance oscillators. In other words, the dynamics of the electronic device is governed by the same equations as the TBHO. Therefore, it allows us to experimentally investigate the rich dynamics of the system in the whole space of parameters. Also, using computer simulations the average Lyapunov exponent, regular and chaotic regions can be identified. We observe an excellent agreement between the numerical simulation and the experiment. 

\section{Fundamentals of the model}

\begin{figure}[t]
\includegraphics[width=10.0cm]{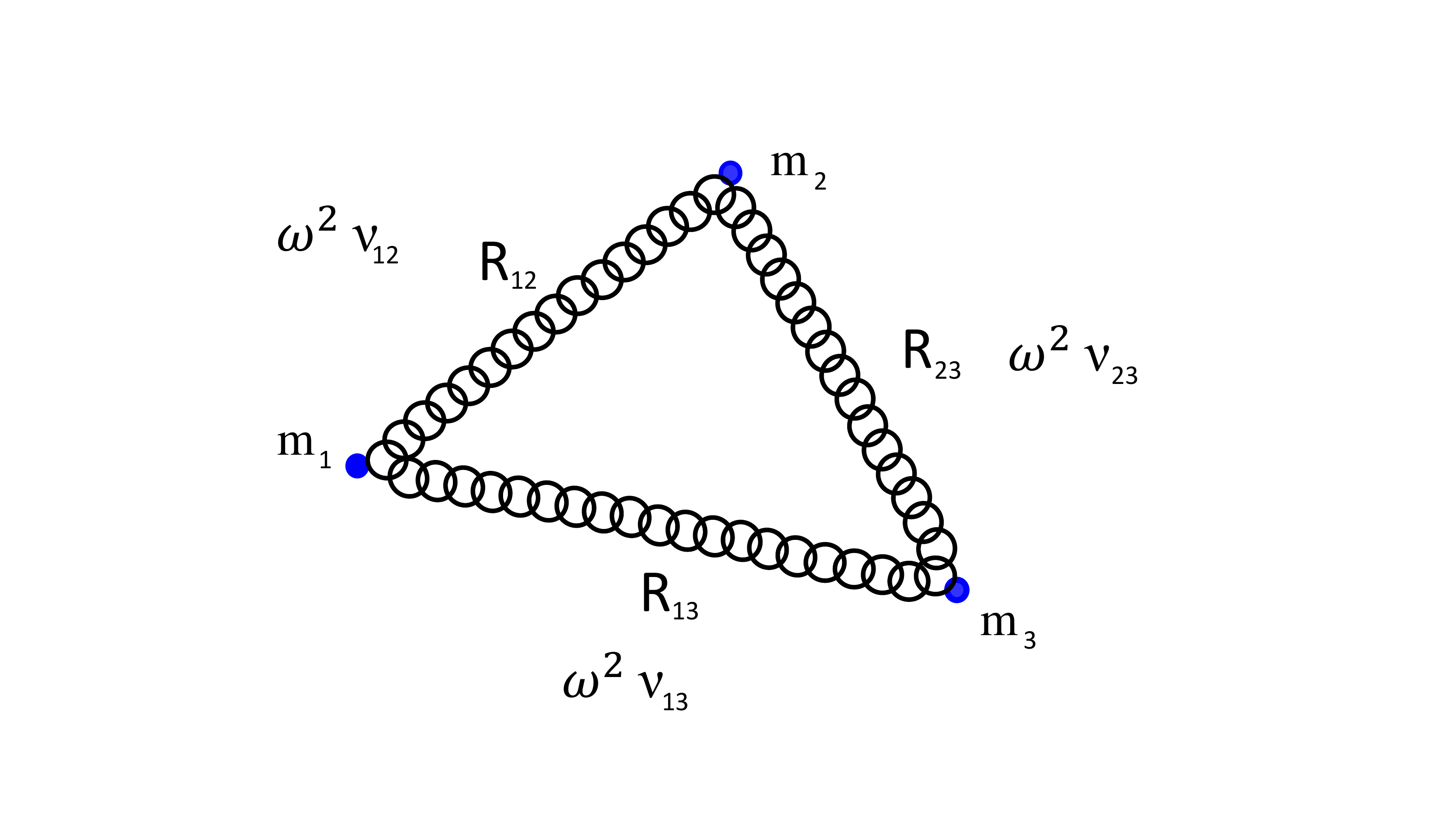}
\caption{The three-body harmonic oscillator (TBHO) in $\mathbb{R}^d$ ($d>1$) with finite rest lengths. The configuration of equilibrium occurs when the distances between the bodies equal
the rest lengths, $r_{ij}=R_{ij}$, and the bodies are at rest.}
\label{Geometrical settings}
\end{figure}

Let us consider in $\mathbb{R}^d$ ($d>1$) the TBHO at zero total angular momentum with the center of mass chosen as the origin of the reference frame. The corresponding Hamiltonian of the collision- and collinear configuration-free TBHO is of the form \cite{AMER2021},\cite{Murnaghan}
\begin{equation}
\label{Hhar}
\begin{aligned}
 {H}\  & =  \
\frac{1}{2}\,\bigg[\,\frac{P_{12}^2}{m_{12}} \ + \ \frac{P_{13}^2}{m_{13}} \ + \ \frac{P_{23}^2}{m_{23}}
\ + \ 
\\ & 
 \frac{r^2_{12} + r_{13}^2- r^2_{23}}{r_{12}\,r_{13}\,m_1}\,P_{12}\,P_{13}
\ + \
 \frac{r^2_{12} + r^2_{23}- r^2_{13}}{r_{12}\,r_{23}\,m_2}\,P_{12}\,P_{23}
\\ &
\ + \   \frac{r^2_{23} + r^2_{13}- r^2_{12}}{r_{23}\,r_{13}\,m_3}\,P_{23}\,P_{13}\,\bigg]\  + \ V(r_{12},\,r_{13},\,r_{23}) \ ,
\end{aligned}
\end{equation}
here $m_{ij}\equiv \frac{m_i\,m_j}{m_i+m_j}$ are reduced masses ($i\neq j=1,2,3$), $r_{ij}={\mid {\bf r}_i-{\bf r}_j\mid}$ denote the relative distances, $P_{ij}$ the associated canonical momentum variables, and
\begin{equation}
\label{V3-es}
\begin{aligned}
   V \ & = \    2\,\omega^2\big[   \nu_{12}\,{({r_{12}}\,-\,R_{12})}^2 \ + \  \nu_{13}\,{({r_{13}}\,-\,R_{13})}^2 
   \\ & 
   \ + \ \nu_{23}\,{({r_{23}}\,-\,R_{23})}^2\, \big] \ ,
\end{aligned}
\end{equation}
where $\omega > 0$ is the frequency, $R_{ij}\geq 0$ are the rest lengths of the system and $\nu_{ij}$ define spring constants, see Fig. \ref{Geometrical settings}. 

\bigskip

\begin{figure*}[t]
\includegraphics[width=18cm]{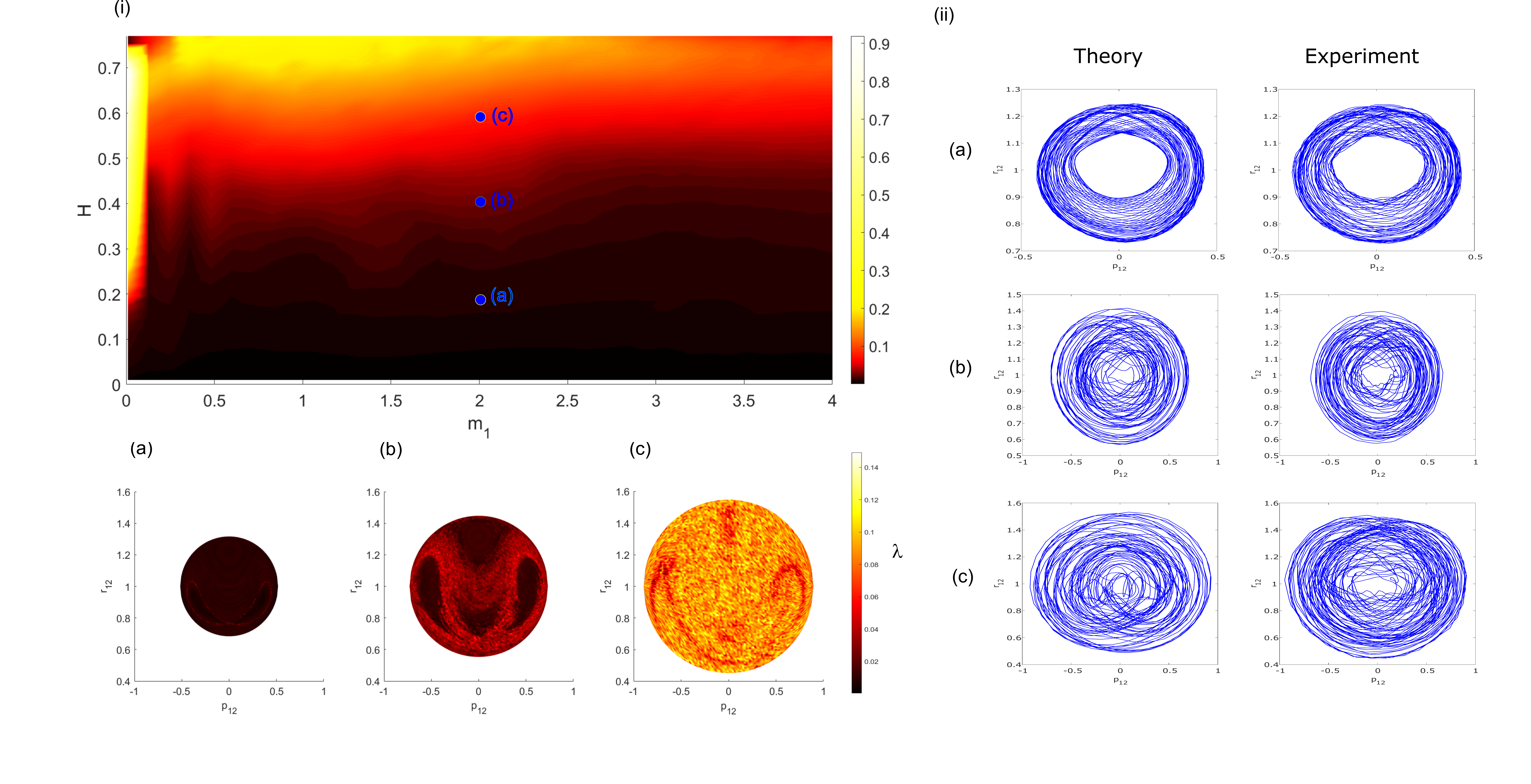}
\caption{(i) Map of the average largest Lyapunov exponents as a function of the energy $H$ and the mass $m_1$. In the bottom panels show maps of largest Lyapunov exponents in the plane ($P_{12}$, $r_{12}$), for three different energies: (a) $H=0.2$, (b) $H=0.4$ and (c) $H=0.6$, considereing $m_1=2$. Note that each pair $(H,m_1)$ is marked with blue dots in the main map. ii) Projections of the trajectories in the plane $(P_{12},r_{12})$ for numerical simulations (left-hand side) and experimental results (right-hand side), corresponding to the initial conditions $(P_{12},P_{31},P_{23},r_{12},P_{31},r_{23})$: (a) (-0.38077,0,0,0.91153,1.19442,1), (b) (-0.68603,0,0,0.94579,1.14342,1) and (c)(-0.60531,0,0,1.37879,1.13824,1).} 
\label{fig:lle}
\end{figure*}

The Hamiltonian (\ref{Hhar}) is the symmetric reduction of the original $3d-$dimensional problem. It governs all the trajectories of the original problem possessing zero total angular momentum. This is valid for any pairwise potential $V=V(r_{12},r_{13},r_{23})$. Interestingly, it can also be interpreted as the Hamiltonian for a three-dimensional moving body in a curved space \cite{AMER2021}. 

Alternatively, one can promote the mutual distances (squared), namely $\rho_{ij}=r_{ij}^2$, as new dynamical coordinates complemented by their corresponding canonical momentum variables. The benefits of this representation yield in the fact that (i) it encodes the discrete symmetries of the original $3d-$dimensional kinetic energy, (ii) the kinetic energy in the reduced Hamiltonian is simply a polynomial expression and (iii) it nicely establishes an interesting link between the dynamics of the TBHO and the geometrical quantities of the so called \emph{triangle of interaction}, the triangle formed by the vector positions of the three particles. 

The corresponding co-metric in (\ref{Hhar}) is given by
\begin{equation}
\label{gmn33-rho}
 g^{\mu \nu}\ = \frac{1}{2}\left(
 \begin{array}{ccc}
 \frac{1}{m_{12}}  & \, \frac{(r^2_{13} + r^2_{12} - r^2_{23})}{2\,r_{13}\,r_{12}\,m_1} & \, \frac{(r^2_{23} + r^2_{12} - r^2_{13})}{2\,r_{23}\,r_{12}\,m_2} \\
            &                                   &                                   \\
 \frac{(r^2_{13} + r^2_{12} - r^2_{23})}{2\,r_{13}\,r_{12}\,m_1} & \,  \frac{1}{m_{13}}  & \, \frac{(r^2_{13} + r^2_{23} - r^2_{12})}{2\,r_{13}\,r_{23}\,m_3} \\
            &  \                                  &                                   \\
\frac{(r^2_{23} + r^2_{12} - r^2_{13})}{2\,r_{23}\,r_{12}\,m_2} & \,  \frac{(r^2_{13} + r^2_{23} - r^2_{12})}{2\,r_{13}\,r_{23}\,m_3} & \frac{1}{m_{23}} 
 \end{array}
               \right) \, , \nonumber
\end{equation}
and its determinant
\begin{equation}
\label{gmn33-rho-det-M}
 \mid g \mid \ \equiv \ {\rm Det} g^{\mu \nu}\ = \ \frac{M^2}{2\,m_1^2m_2^2m_3^2} \,\frac{{\cal I}\,S_{\triangle}^2}{r_{12}^2\,r_{13}^2\,r_{23}^2\,} \ ,
\end{equation}
($M=m_1+m_2+m_3$) factorizes. The quantities 
\[
  S_{\triangle}^2 =\frac{  2r^2_{12}\,r^2_{13}+2r^2_{12}\,r^2_{23}+2r^2_{23}\,r^2_{13}-r_{12}^4-
  r_{13}^4-r_{23}^4}{16} \, ,
\]
\[
{\cal I} \ = \ \frac{m_1 \,m_2\, r^2_{12}\ + \ m_1\, m_3\, r^2_{13}\  +\ m_2 \,m_3 \,r^2_{23}}{M} \ ,
\]
possess a clear geometrical interpretation. The term $S_{\triangle}^2$ is the area squared of the triangle of interaction whilst ${\cal I}$ is the moment of inertia of the system. In the case of 3 equal masses, one can further introduce the 3 lowest elementary symmetric polynomials $\sigma_i \ (i=1,2,3)$ in $\rho-$coordinates as generalized coordinates. They are invariant under the action of $\mathbb{Z}_2^3 \oplus \mathcal{S}_3^2$, reflections ($r_{ij} \rightarrow -r_{ij}$) and permutations of the $\rho$-variables. In this representation, the associated reduced kinetic energy is a polynomial again. In particular, at $d=3$ the discriminant of the 4th degree polynomial equation, solution of which corresponds to the physically relevant 3-body Newtonian potential in terms of  $\sigma-$coordinates, do coincide with one of the factors of the determinant associated to the corresponding co-metric\cite{AMER2021}.

\bigskip
\begin{figure}[t]
\includegraphics[width=8.5cm]{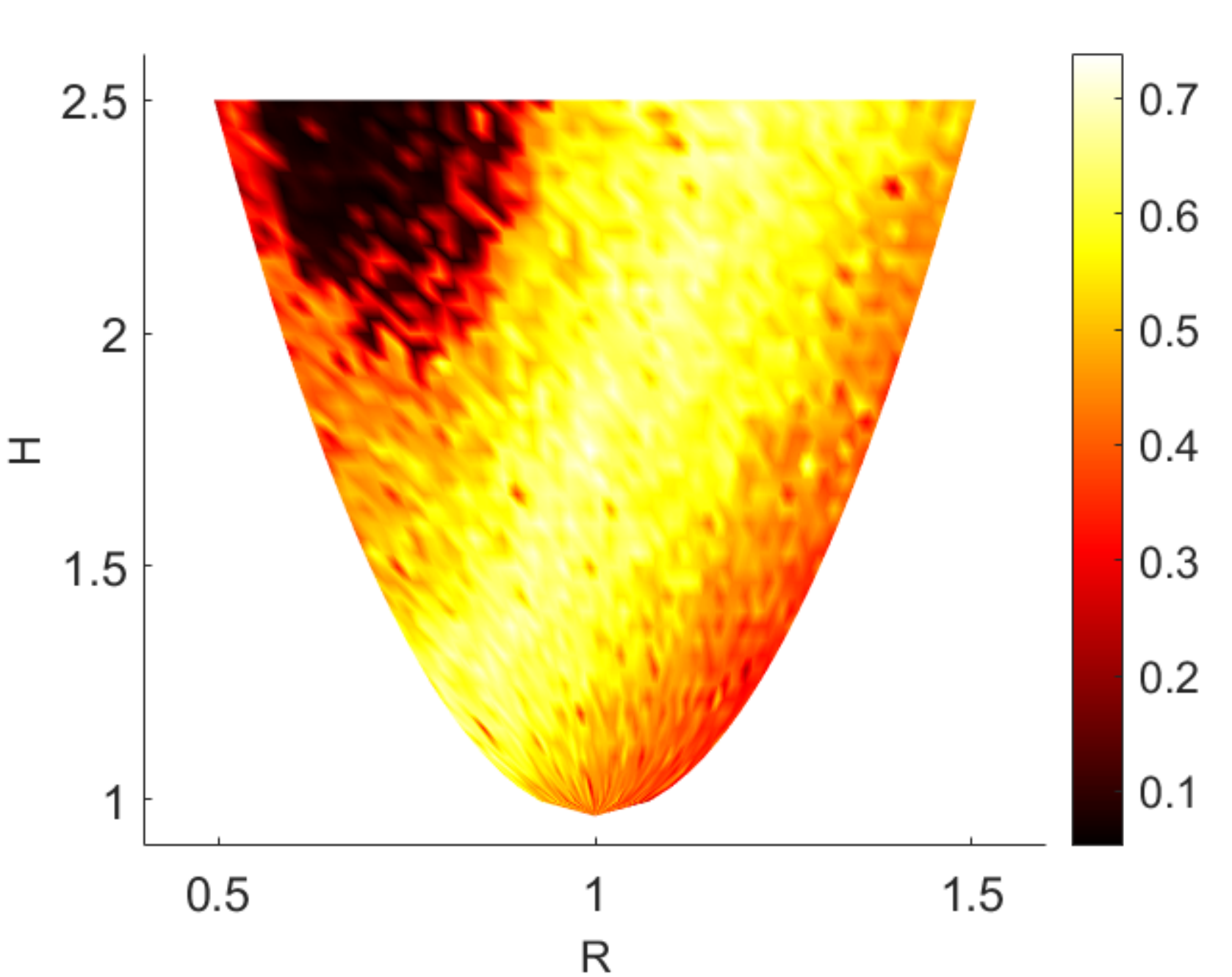}
\caption{Map of largest Lyapunov exponents as a function of the energy $H$ and the equilibrium distance $R$. For each pair ($H$,$R$) the Lyapunov exponent is computed by integrating the motion equations of the system in cartesian coordinates in the restricted phase space defined by $H$, $x_{1}=0$, $x_{2}=-1/2$, $y_{1}=1/ \sqrt{3}$, $y_{2}=-1/ 2\sqrt{3}$, $p_{x_2}=-1/2$ and $p_{y_1}=-1$.}
\label{figHvsR}
\end{figure}

\bigskip
\begin{figure}[t]
\includegraphics[width=8.5cm]{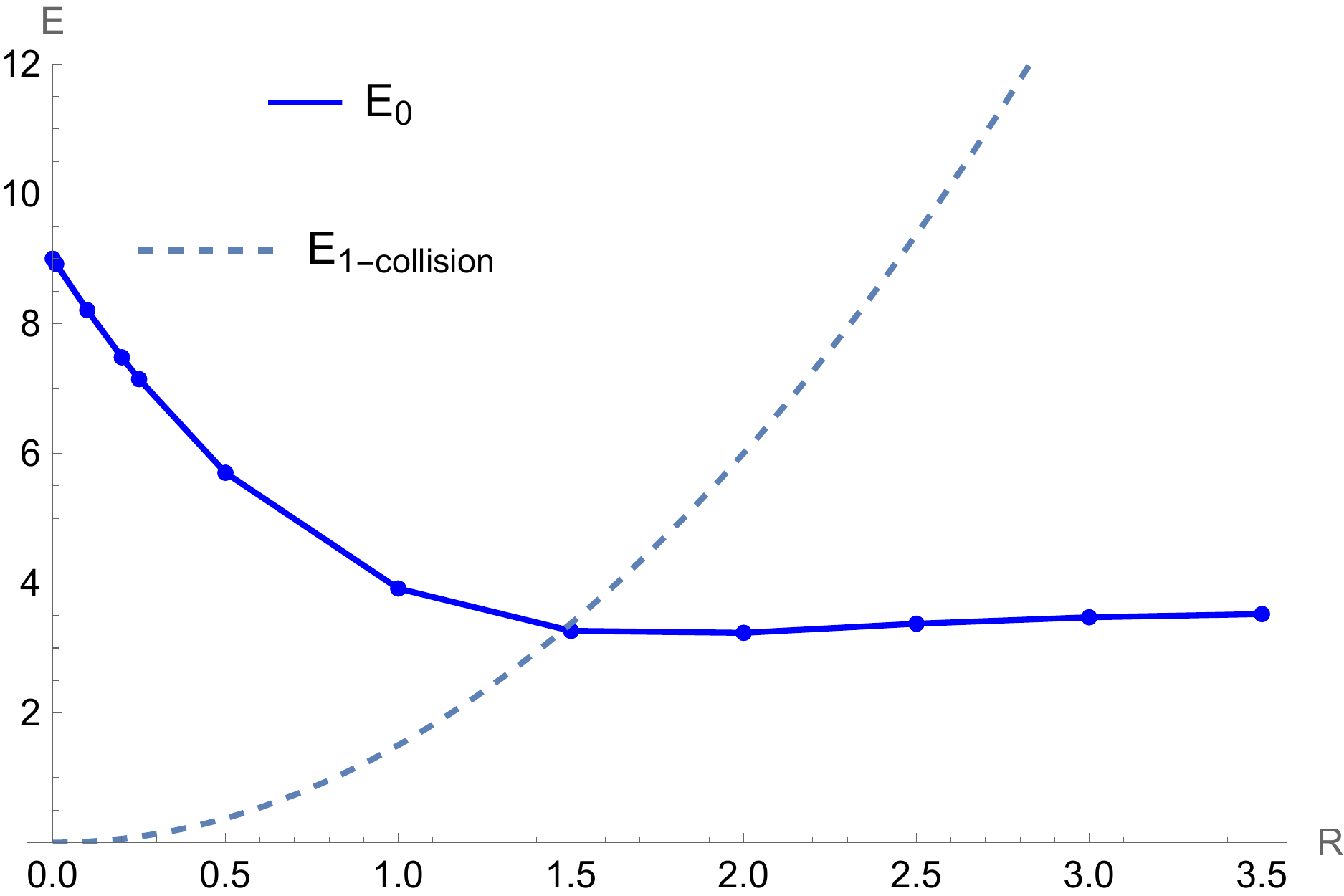}
\caption{Ground state energy $E_0$(a.u.) of the quantum TBHO \textit{vs} $R$(a.u.) at $d=3$. The solid line corresponds to an interpolation of the numerical results marked by dots whilst the dashed line refers to the classical critical value $E_{1-collision}$ below which no collisions occur. The parameters $m_1=m_2=m_3=1$, $\nu_{12}=\nu_{13}=\nu_{23}=\frac{3}{4}$ and $\omega=1$ were used. }
\label{E0vsR}
\end{figure}

\section{Chaotic and periodic regimes}

\begin{figure*}[t]
\includegraphics[width=18cm]{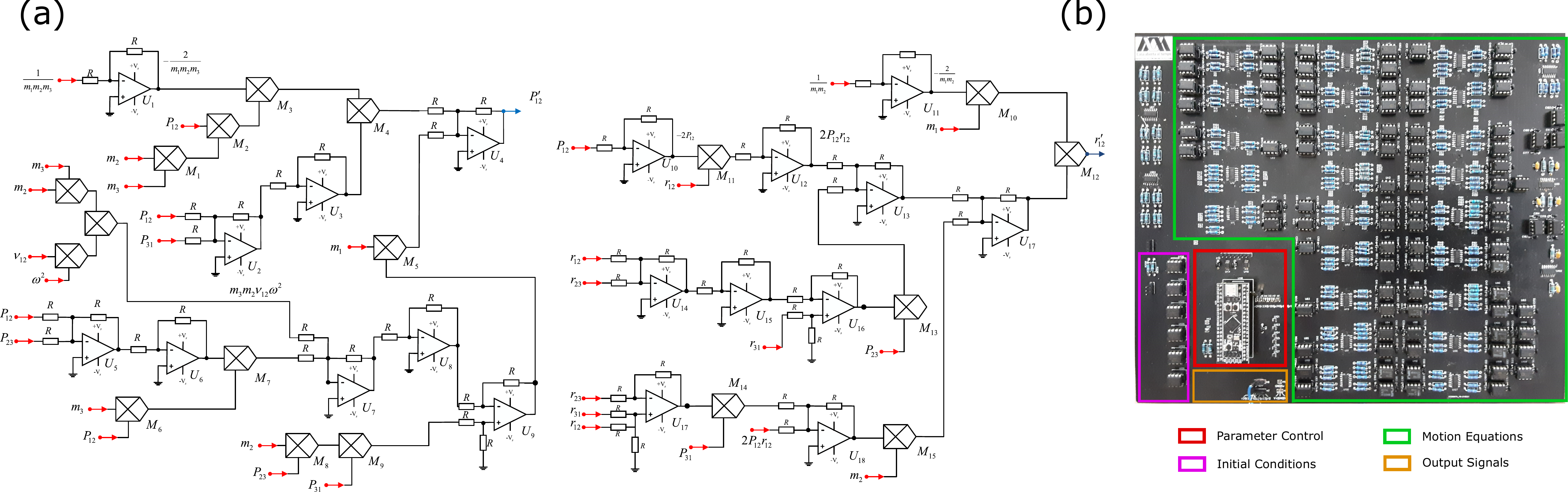}
\caption{(a) Scheme of the electronic circuit for the canonical variables ${P}_{12}$ and ${r}_{12}$. Our experimental setup consist of linear and nonlinear configurations of operational amplifiers (OPAMPs) interconnected properly to satisfy the motion equations obtained from the Hamiltonian (\ref{Hhar}). These configurations implement resistors ($R_j$), capacitors ($C_j$), general-purpose OPAMPs ($U_j$) and analog multipliers ($M_j$). Here, the red and blue nodes denote input and output signals, respectively. (b) Printed circuit board of the three-body chain of harmonic oscillators. The PCB was designed in EasyEDA Software.}
\label{fig:circuit}
\end{figure*}

We perform simulations by numerically integrating the equations of motion obtained from Hamiltonian (\ref{Hhar}), in order to characterize periodic and chaotic regions for different values of the energy. We study the nature of the trajectories by computing the largest Lyapunov exponents. The average largest Lyapunov exponents as a function of the energy $H$ and the mass $m_1$ are shown in Figure \ref{fig:lle}(i). We compute the average Lyapunov exponent for each pair ($H$,$m_1$) by integrating the motion equations for 2500 initial conditions in the state space defined by $H$, $P_{13}=0$, $P_{23}=0$ and $r_{23}=1$. Because of Eq. \ref{gmn33-rho-det-M}, the dynamics associated with configurations for which the area of the triangle of interaction, during time evolution, vanishes are not considered to calculate the average largest Lyapunov exponents. Here, the masses and spring constants are set to $m_2=m_3=1$, $\nu_{12}=\nu_{31}=\nu_{23}=1$, $\omega=1$ and $R_{12}=R_{13}=R_{23}\equiv R= 1$. We selected four different energies, whose values are indicated on the phase map by blue points: (a) $H=0.2$, (b) $H=0.4$ and (c) $H=0.6$. Detailed maps of the largest Lyapunov exponents in the plane ($P_{12}$, $r_{12}$), corresponding to the selected energies are shown on the lower panel of Fig. \ref{fig:lle}. For low energies, regular dynamics can be found only ($H=0.2$). For energies close to $H =0.4$, mixed zones where chaos and regularity coexist acquire relevance in the dynamics. When the energy increases up to $H = 0.6$, chaotic dynamics predominate. 

To further illustrate the rich dynamic of the system, in Fig.\ref{figHvsR} the nontrivial dependence of the largest Lyapunov exponent as a function of the energy $H$ and the equilibrium distance $R$ is displayed, whereas in the Fig. \ref{E0vsR} the ground state energy $E_0$ of the quantum system calculated with the Lagrange-mesh method \cite{Baye2015TheLM} is presented. In Fig. \ref{figHvsR}, at fixed energy $H=E$ the largest Lyapunov exponent is not a monotonic function of the rest length $R$. Interestingly, in the quantum case we observe from Fig.\ref{E0vsR} that a local shallow minimum of $E_0(R)$ appears.

Importantly, the three-body chain of harmonic oscillators can be electronically implemented using the working principle of an analog computer. In this device, the physical quantities from the problem to be solved such as fluid, pressure, or velocity, are represented by voltage signals, so the temporal evolution of the variables of interest is directly measured on the analog circuit. This has enabled the implementation of experimental systems that offer certain advantages over the original physical systems \cite{leon2014generation, lamata2018digital, quiroz2020experimental, quiroz2021reconfigurable, jimenez2021experimental}. Especially, analog computers are suited to simulate dynamical systems \cite{vazquez2013arbitrary,noordergraaf1963use, moore1998dynamical, leon2018observation, quiroz2019generation, quiroz2021demand}, and even, stochastic dynamical systems \cite{leon2015noise, quiroz2017emergence}. A potential advantage of analog computers is the simulation of continuous physical quantities by performing real-time simultaneous operations. Particularly, when it is sought to integrate dynamical systems that include high-frequency components, digital computers demand high computational resources because they require small steps during numeral integration whereas analog computers carry out a continuous integration. Here, we exploit this fact to build an electronic version of the three-body chain of harmonic oscillators and to experimentally investigate the presence of regular and chaotic behaviors on the analog model. The projections of the trajectories in the planes ($P_{12}$, $r_{12}$), for the numerical simulations and experimental results are shown in Fig. \ref{fig:lle}(ii)(a)-(c). For energy $H \sim $ 0.2, the trajectories are periodic, characterized by a Lyapunov exponent close to zero. Note that as energy increases up to $H \sim$ 0.4,  a chaotic trajectory is observed. As Fig \ref{fig:lle}(ii)(c) shows, the phase plane exhibits chaotic dynamic for $H \sim$ 0.7. Note that the experimental results are in quantitative and qualitative agreement with the numerical simulations.

\section{Electronic implementation.} 

We design and implement an experimental setup based on active electrical networks and passive linear electrical components to reproduce the dynamics described by the Hamiltonian (\ref{Hhar}). The active networks consist of interconnecting passive electronic components, and operational amplifiers (OPAMPs) to satisfy basic configurations whose transfer functions are equivalent to mathematical operations, namely, addition, subtraction, integration, and amplification. These electronic circuits are typically employed to implement linear operations; however, it is possible to carry out nonlinear functions, such as products and divisions between signals from logarithm and antilogarithm configurations connected in cascade or parallel. Both functions comprise linear circuits in which the output voltage is proportional to the natural logarithm and antilogarithm of the input, respectively. Interestingly, nonlinear operations are typically performed through integrated analog multipliers. The main reason for this is that log and anti-log amplifiers include diodes and transistors in their feedback, which leads to a high dependence on the temperature. In addition, the bandwidth is limited by the  signal amplitude. It is worth mentioning that through connecting linear and nonlinear basic configurations in a structured way, it is possible to build complex operations where the involved voltage signals represent physical variables of the system of interest. We perform nonlinear functions between signals such as multiplication/division and squaring by employing analog multipliers AD633JN. In the analog multipliers, the manufacturer includes a factor that divides the output voltage over ten, in order to avoid saturation of the device.

Traditionally, the parameters of the physical system are mapped in passive components on the electronic circuit, specifically, in the relationship of two resistors or resistor and capacitor.  Under this condition, the values that can take the parameters are limited to the commercial values of the components. To overcome this limitation, we introduce the system parameters ($m_1$, $m_2$, $m_3$, $\nu_{12}$, $\nu_{23}$, $\nu_{31}$, $\omega$, $R$) through voltage signals generated by digital-to-analog converters, which are manufactured by Microchip with series MCP4921. These devices can take discrete values between 0V and 5V with a resolution of 1.22 mV (12 bits) by allowing to select parameters from a broad range of possible values. Note that we have transferred the issue from replacing passive electronic components to carrying out the product of voltage signals with analog multipliers.

Because of their importance, the initial conditions [$P_{12}(0)$, $P_{23}(0)$, $P_{13}(0)$, $\rho_{12}(0)$, $\rho_{23}(0)$, $\rho_{31}(0)$] are also set via voltage values. In this sense, each of the voltage signals is individually controlled via software. Using a serial peripheral interface (SPI) protocol, digital-to-analog converters are communicated to a master system board, which incorporates a 32-bits STM32F401CC microcontroller from the ARM Cortex-M4 family.  Fig. \ref{fig:circuit}a shows the scheme of the designed electronic circuit. Because canonical variables ($P_{ij},r_{ij}$) of the Hamiltonian (\ref{Hhar}) are described by motion equations with a similar structure, we show solely the electronic circuit for the variables $\dot{P}_{12}$ and $\dot{r}_{12}$. There, $R_j$ , $C_j$, $U_j$, and $M_j$ stand for resistors and capacitors, general-purpose operational amplifiers LF353, and analog multiplier AD633JN, respectively. Here the input and output signals are indicated with red and blue nodes, individually.

\begin{figure}[t]
\includegraphics[width=6cm]{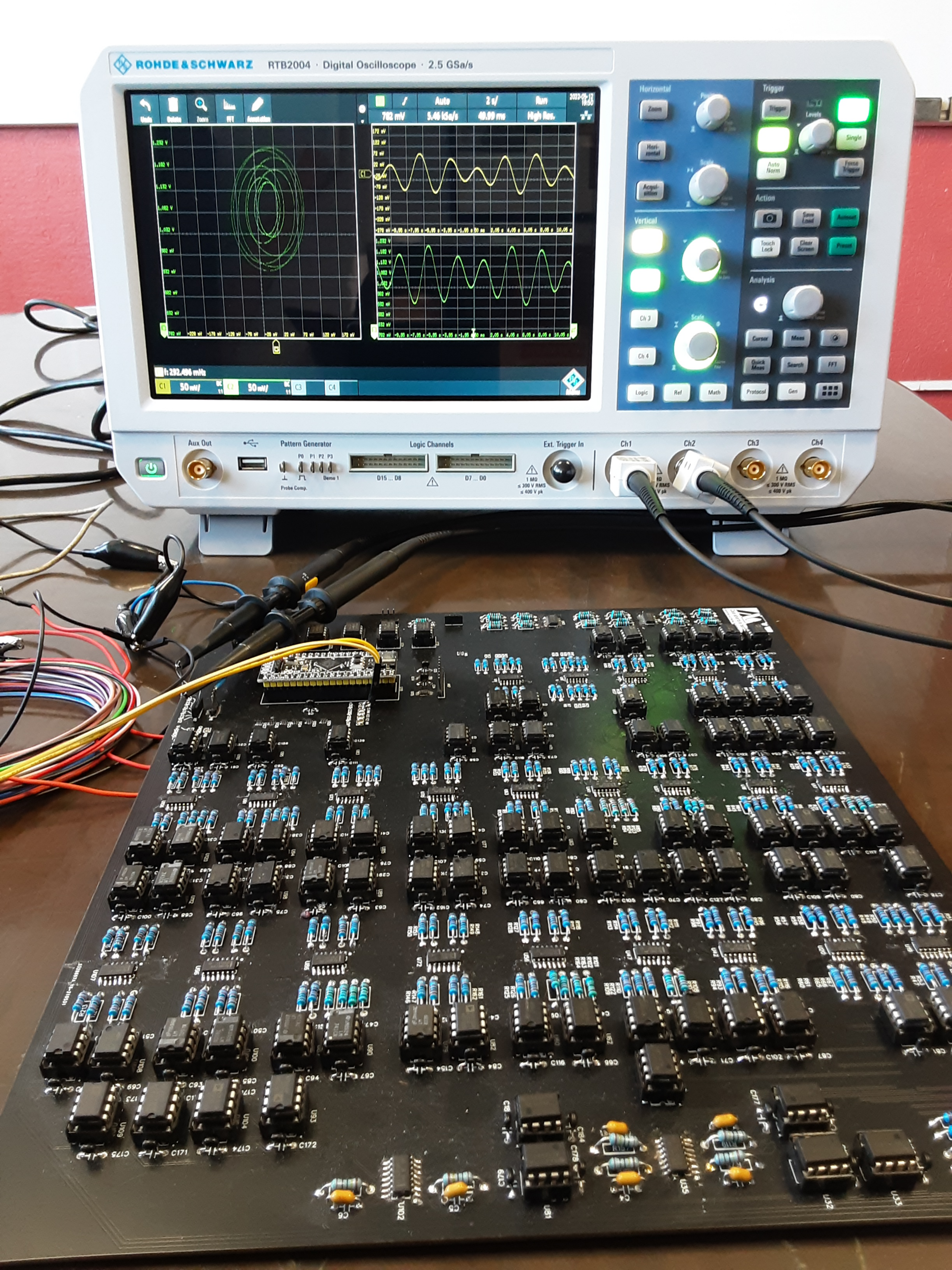}
\caption{Experimental setup of the classical harmonic three-body system.}
\label{setup}
\end{figure}

Figure \ref{fig:circuit}b shows the printed circuit board of the classical three-body closed chain of harmonic oscillators. Electrical contact failures were avoided by mounting and soldering all electronic components on a printed circuit board (PCB, 20x20 cm). In Figure \ref{fig:circuit}b, the initial conditions and control parameter modules are denoted with magenta and red squares, respectively.  Note that the motion equations occupy a great part of the PCB part (green square).  A stabilized DC power supply (KEITHLEY Triple channel, 2231A-30-3) was used to energize our setup, whereas the acquisition of the output signals (orange square), corresponding to the physical variables of the system, was performed with Rohde-Schwarz oscilloscope. The experimental setup mounted in the laboratory is shown in Fig. \ref{setup}.

\bigskip

\section{Conclusions}
In summary, we have presented a novel electronic experimental realization of the classical three-body closed chain of harmonic oscillators. Our setup allowed us to explore different regions of behavior due to the fact that the intrinsic parameters and initial states of the system are independently set by voltage inputs. Complementary phase planes and Lyapunov exponents were computed to characterize periodic and chaotic regions for different values of the energy. Importantly, the experimental results are in quantitative and qualitative agreement with the numerical simulations.

\begin{acknowledgements}
The Authors are grateful to L. Jiménez-Lara for their interest in the work and useful discussions. This work is partially supported by  Programa Especial de Apoyo a la Investigación 2021, UAM-I.  
\end{acknowledgements}

\appendix
\section{Scaling mass relation}
One can also use the variables $(L_1,\,L_2,\,\sigma)$
\begin{equation}\label{}
\begin{aligned}
& r_{12}^2 \ = \ (\, L_2^2 \ - \ L_1^2 \,)\,\sin \sigma\,\cos \sigma \\
& r_{13}^2 \ = \ L_2^2\,\sin^2 \sigma \ + \ L_1^2\,\cos^2 \sigma  \\
& r_{23}^2 \ = \ L_2^2\,\cos^2 \sigma \ + \ L_1^2\,\sin^2 \sigma \ ,
\end{aligned}
\end{equation}
(see Ref.\cite{Pina} and references therein) which are useful in the study of the general three-body problem. In these variables, the potential (\ref{V3-es}) takes the form
\begin{equation}
\label{}
\begin{aligned}
   V & \ =\ 2\,\omega^2\bigg[   \nu_{12}\,{\bigg(\,\sqrt{(\, L_2^2 \ - \ L_1^2 \,)\,\sin \sigma\,\cos \sigma}\,-\,R_{12}\,\bigg)}^2
\\ &
   \ + \  \nu_{13}\,{\bigg(\,\sqrt{L_2^2\,\sin^2 \sigma
    \ + \ L_1^2\,\cos^2 \sigma }\,-\,R_{13}\,\bigg)}^2 \ + \
\\ &    \nu_{23}\,{\bigg(\,\sqrt{L_2^2\,\cos^2 \sigma \ + \ L_1^2\,\sin^2 \sigma}\,-\,R_{23}\,\bigg)}^2 \, \bigg]  \ .
\end{aligned}
\end{equation}
Accordingly, the reduced Hamiltonian (\ref{Hhar}) becomes
\begin{equation}
\label{}
  {\cal H} \   = \ \frac{1}{2\,{\cal M}} \bigg(\, {P}_{L_1}^2 \ + \ {P}_{L_2}^2 \ + \ \frac{ L_2^2\,+\,L_1^2}{{(L_2^2\,-\,L_1^2)}^2}\,{P}_{\sigma}^2 \,\bigg) \ + \  V \ ,
\end{equation}
here ${\cal M} \equiv \sqrt{\frac{m_1\,m_2\,m_3}{m_1+m_2+m_3}}$. Eventually, the following scaling relation emerges
\begin{equation}\label{}
  {\cal H}\big[\,{\cal M},\,\omega\,\big] \ = \ \frac{\tilde {\cal M}}{{\cal M}}\,{\cal H}\big[\,\tilde {\cal M},\,\omega\,{\cal M}/\tilde {\cal M}\,\big] \ ,
\end{equation}
where $\tilde{\cal M} \equiv \sqrt{\frac{\tilde m_1\,\tilde m_2\,\tilde m_3}{\tilde m_1+\tilde m_2+\tilde m_3}}$. It allows us to connect two systems with different masses. Unlike the $\rho$-representation, the Hamiltonian ${\cal H}$ is not a polynomial function in the variables $(L_1,\,L_2,\,\sigma)$.

\nocite{*}

\bibliography{apssamp}

\end{document}